\documentclass[aps,prb,11pt]{revtex4-2} 
\usepackage{graphicx,bm,amssymb}

\def\be{\begin{equation}}
\def\ee{\end{equation}}
\def\ba{\begin{eqnarray}}
\def\ea{\end{eqnarray}}

\def\bc{\begin{center}}
\def\ec{\end{center}}
\def\p{\partial}

\begin{document}

\title{Strong photoresponse of edge two-dimensional electrons in a  magnetic field }

\author{Sergey A. Mikhailov}
\email[Electronic mail: ]{sergey.mikhailov@physik.uni-augsburg.de} 
\affiliation{Institute of Physics, University of Augsburg, D-86135 Augsburg, Germany} 
\author{Wladislaw Michailow} 
\email[Electronic mail: ]{wladislaw.michailow@swansea.ac.uk}%{wm297@cam.ac.uk} 
\affiliation{Centre for Integrative Semiconductor Materials and Department of Physics, Swansea University Bay Campus, Swansea, SA1 8EN, United Kingdom} 

\date{27.05.2026}

\begin{abstract}
Electrons in two-dimensional electron gases in the presence of an out-of-plane magnetic field propagate along the edge with a high velocity of the order of the Fermi velocity. Under microwave and terahertz radiation, photon absorption by these electrons provides a pathway to realising sensitive radiation detection. Here, we develop a detailed quantum theory of the photocurrent generated in such a system by an incident electromagnetic wave and propose an experimental geometry for observing the predicted phenomenon. By using suitably arranged radiation confinement structures, a strong photocurrent generation efficiency can be obtained. We also demonstrate that the resulting photoresponse in III-V semiconductor structures can be orders of magnitude higher than that measured in graphene.
\end{abstract}

\maketitle

\section{Introduction\label{sec:intro}}

The interaction of two-dimensional (2D) electron systems with microwave and far-infrared electromagnetic radiation has always attracted great interest from researchers because of its fundamental importance and potential for useful electronic applications. As early as 1967, Stern \cite{Stern67} calculated the spectrum of plasma waves in a 2D electron gas (2DEG) and showed that, unlike plasma oscillations in 3D metals, the frequency of which lies in the ultraviolet range and cannot be easily changed by external parameters, the frequency $\omega_p$ of 2D plasmons depends on the electron density $n_s$, the plasmon wave vector $q$, and the effective dielectric permittivity of the surrounding medium $\epsilon$,
\be 
\omega_p^2(q)=\frac{2\pi n_se^2}{m\epsilon}q,
\label{2Dplasmon}
\ee
where $e$ and $m$ are the charge and effective mass of 2D electrons. Depending on the parameters ($n_s$, $q$, $\epsilon$, $m$), the frequency of the 2D plasmons can be tuned in a very wide and technologically important range from microwaves to terahertz.

The first experiments on the detection of 2D plasmons (\ref{2Dplasmon}) were carried out using the far-infrared transmission spectroscopy \cite{Allen77,Theis78,Theis80,Heitmann86}. In such experiments, a semiconductor structure with a 2DEG, e.g., a Si-MOSFET (metal-oxide-semiconductor field-effect transistor), was covered by a semitransparent metallic gate and a periodic metallic grating. The grating period $a$ determined the wave vector $q\simeq 2\pi/a$ of the 2D plasmons, while the electron density $n_s$ could be controlled by applying a dc voltage between the 2DEG and the gate. The measured far-infrared transmission spectra showed resonances at the frequency (\ref{2Dplasmon}) with the linewidth $\delta\omega_p\simeq 1/\tau_p$ determined by the Drude momentum relaxation time $\tau_p$. 

Since the beginning of the 2D plasmon research attempts have been made to exploit the 2D electron systems for applications in microwave and THz electronics. In 1980, Tsui with co-authors \cite{Tsui80} attempted to realize Smith-Purcell-type  emission from a 2D electron system with a grating, see also \cite{Hirakawa95}. They passed a strong dc electric current between two (source and drain) contacts to the 2DEG which caused the electrons to move with a drift velocity $v_{\rm dr}$. The system was expected to emit radiation at a frequency $f\simeq v_{\rm dr}/a$, similar to that observed in free-electron lasers or vacuum backward-wave oscillators. However, weak thermal radiation at the frequency of the 2D plasmons (\ref{2Dplasmon}) was observed instead. A detailed theory explaining under what conditions Smith-Purcell radiation should be observed, and under what conditions thermal radiation at the frequency of 2D plasmons, was developed in Ref. \cite{Mikhailov98c}.

The use of 2D plasmons for detecting sub-THz radiation has been more successful. In 1996, Dyakonov and Shur \cite{Dyakonov96} proposed using the excitation of plasma waves in 2D electron systems to detect electromagnetic radiation. They considered a 2DEG-structure that has source and drain contacts and is covered by a metal gate. It was assumed that a narrow gap between the gate and the source is illuminated by focused electromagnetic radiation. This focused radiation generates plasma waves in the 2DEG, propagating from the source to the drain. Due to the nonlinearities of the classical hydrodynamic equations that describe this process, a time-independent photovoltage is generated between source and drain \cite{Dyakonov96}, -- the detection mechanism is nowadays referred to as plasmonic mixing. Later, other geometries such as an open plasmonic cavity with a narrow gate partly covering the 2DEG \cite{Popov2005} or an asymmetric double-grating-gate FET structure \cite{Popov2011} have been also proposed for plasmonic detection of sub-THz and THz radiation.

Experimental work on semiconductor 2DEG structures (e.g. Refs. \cite{Lu01,Knap02,Sun2012,Kurita14,Giliberti15,Sun19,Sun20,Satou23,Yadav24}) and on graphene (e.g. Refs. \cite{Olbrich16,Moench22,Delgado22,Abidi24}) confirmed that the photoresponse is indeed observed in the geometries proposed in the theoretical papers \cite{Dyakonov96,Popov2005,Popov2011}. Currently, such observations are usually interpreted as a consequence of the plasmonic mixing effect. However, a quantitative comparison of the experimentally observed and the theoretically predicted photoresponse, carried out in Ref. \cite{Kang06}, showed that the experimentally observed photovoltage can be significantly (approximately 4 times) higher than that predicted by the plasmonic mixing theory. This evidently suggested that, in addition to the mechanism predicted in \cite{Dyakonov96}, there must be another effect that leads to the occurrence of a dc photoresponse in a gated 2DEG under illumination.

With the discovery of graphene, in which a similar phenomenon was also observed, it was proposed that the photothermoelectric (PTE) effect is responsible for the enhanced  photoresponse in graphene-based FETs \cite{Gabor11}. However, graphene is a quite specific material. The energy dispersion of electrons in it is linear, with the Fermi velocity $\sim 10^8$ cm/s almost three orders of magnitude higher than that of acoustic phonons in typical semiconductors ($\sim 5\times 10^5$ cm/s). This greatly hinders the energy exchange between the electrons and the surrounding crystal lattice and contributes to significant heating of the electron gas in graphene. In conventional semiconductors with parabolic energy dispersion, electron heating is much less efficient, as the electrons easily transfer their energy to the crystal lattice, emitting acoustic phonons. Therefore, the PTE effect cannot explain the observed difference between the experimental results and the plasmonic mixing prediction in conventional semiconductor systems. In some papers, e.g., in \cite{Regensburger24}, the photoresponse of FETs based on conventional semiconductors was attributed to the PTE effect. However, a quantitative analysis of the expected strength of the PTE response was lacking, nor was it considered whether other mechanisms, such as the in-plane photoelectric effect (IPPE) \cite{Michailow22} that we introduce below, could be a possible explanation for the observed response. Moreover, a detailed modeling showed that the PTE effect in 2DEGs based on III-V semiconductors such as AlGaN/GaN is negligible compared to plasmonic mixing \cite{BauerPhD2017}. The question of a possible additional contribution to the photoresponse of field-effect transistors on conventional semiconductors remained open until 2022.

\begin{figure}[ht!]
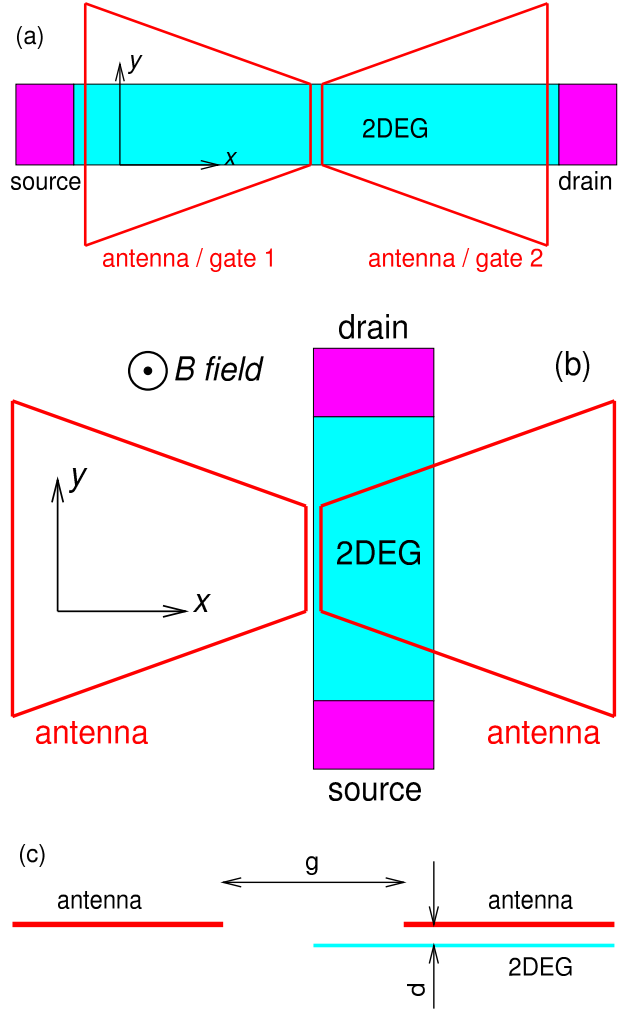

\includegraphics[width=0.49\textwidth]{fig1a.eps}\\ \vspace{5mm}
\includegraphics[width=0.49\textwidth]{fig1b.eps}\\ \vspace{5mm}
\includegraphics[width=0.49\textwidth]{fig1c.eps}\\
\caption{\label{fig:geom}The geometry (a) of the IPPE experiment and (b,c) of the experiment in the magnetic field proposed in this work; (b) -- top view, (c) -- side view.  }
\end{figure}

In 2022, a new quantum phenomenon was experimentally discovered -- the in-plane photoelectric effect (IPPE) \cite{Michailow22}. The geometry of the IPPE experiment is schematically shown in Figure \ref{fig:geom}(a). A 2DEG of a rectangular form, with two (source and drain) contacts is covered by two antenna wings simultaneously serving as gates. By applying different voltages to the left and right gates, a potential step is created for 2D electrons moving from the source to the drain. Without irradiation, the total flow of electrons moving to the left is compensated by the flow of electrons moving to the right. When an electromagnetic wave hits the sample, the antenna focuses the ac electric field into a narrow gap between the two gates. Electrons entering this region absorb a THz photon and overcome the potential barrier created by the different gate voltages. This results in a net electron flow from the region with a higher electron density to the region with a lower electron density. This leads to a photocurrent, which is about one order of magnitude greater than the photocurrent predicted by the plasmonic mixing mechanism under the same conditions \cite{Michailow22}. This finally explained the contradiction found in Ref. \cite{Kang06}. Further details about the IPPE effect, its full theory, as well as comparison of the IPPE photoresponse with other possible detection mechanisms, such as, e.g., bolometric, electron heating, photothermoelectric effect or photon-assisted tunneling,  can be found in Refs. \cite{Michailow22,Mikhailov22,Mikhailov23}.

The particular appeal of 2D electron systems for creating devices operating at frequencies from microwaves to THz is due to the fact that characteristic resonances, such as (\ref{2Dplasmon}), lie in the frequency range of interest and can be tuned in it. But that is not all. By placing a 2DEG sample in a perpendicular magnetic field $B$, one can additionally influence the resonance frequency (\ref{2Dplasmon}). Chiu and Quinn \cite{Chiu74} calculated the spectrum of the 2D magnetoplasmons propagating in the bulk of the 2D electron system,
\be 
\omega_{mp}^2(q)=\omega_c^2+\omega_p^2(q),
\label{2Dmplasmon}
\ee
where $\omega_c=|e|B/mc$ is the cyclotron frequency. The linewidth of the 2D magnetoplasmon modes is also determined by the momentum relaxation rate $1/\tau_p$.

According to Eq. (\ref{2Dmplasmon}), in finite magnetic fields there are no plasma waves at the frequencies below $\omega_c$. However, in 1985 it was theoretically shown that at frequencies $\omega<\omega_c$, there is a special type of plasma waves, the \textit{edge} magnetoplasmons (EMP) \cite{Volkov85bEng,Fetter85}. These waves propagate along the edge of the 2DEG and in high magnetic fields are strongly localized near it. The most remarkable feature of the EMPs is that their damping is much smaller than the damping of the bulk plasmons and magnetoplasmons and falls down when the magnetic field grows \cite{Volkov86bEng,Volkov88Eng}. As a result, while the bulk plasmons (\ref{2Dplasmon}) and magnetoplasmons (\ref{2Dmplasmon}) could only be observed at frequencies $\gtrsim 0.3$ THz due to the restriction $\omega\tau_p\gtrsim 1$, EMPs were experimentally seen in GaAs/AlGaAs heterostructures at frequencies as low as $\simeq 1$ GHz \cite{Volkov86bEng}. 

The influence of the magnetic field on plasma resonances in 2D electron systems can be also used to expand the possibilities of detecting electromagnetic radiation. In the paper \cite{Vasiliadou93}, the bulk magnetoplasmon resonances were observed in the experiments on microwave photoconductivity. The observed magnetoplasmon resonance could be shifted by the magnetic field and the geometric parameters of the sample, offering the possibility of tunable resonant detection of microwave radiation. After the GaAs/AlGaAs samples with extremely high electron mobility ($\mu>10^7$ cm$^2$/Vs) were obtained, a new type of photoresistance oscillations which were called microwave-induced zero resistance states were experimentally observed \cite{Mani02,Zudov03}, see also \cite{Zudov01}. In these experiments, giant oscillations of the microwave-induced resistance of the 2DEG, associated with the harmonics of the cyclotron resonance, were observed. Although this phenomenon is extremely interesting from the fundamental point of view, it can hardly be used to design photodetectors, because the effect is observed only in samples with extremely high electron mobility and at very low temperatures. 
But at about the same time, other photoresistance and photovoltage oscillations related to the excitation of EMPs in the 2DEG were experimentally discovered in Ref. \cite{Kukushkin04a}. They are periodic in the magnetic field $B$ and can be directly used not only for detection, but also for spectroscopy of microwave radiation \cite{Kukushkin05a}. Moreover, such oscillations persist up to high temperatures, at least up to $T\sim 100$ K \cite{Kukushkin05a}. 

As seen from this brief overview, 2D electron systems, both in zero and finite magnetic fields, offer wide possibilities for detecting microwave to THz radiation.

In this paper, we propose a new method of detecting THz radiation in a 2D electron system placed in a magnetic field, and develop a quantum theory of the proposed effect. Like in the IPPE effect, the generation of a dc photocurrent is caused by an absorption of photon quanta in the narrow gap between two wings of an antenna, see Fig. \ref{fig:geom}(b). In contrast to the IPPE effect, the photoexcited electrons do not need to overcome a potential barrier, but exit the irradiated region in a perpendicular direction along the edge of the 2DEG.

Let us briefly explain the physics of the discussed phenomenon. Let us consider a 2DEG in the form of a Hall bar, with source and drain contacts, placed in a perpendicular magnetic field $B$, Fig. \ref{fig:geom}(b). If the magnetic field is sufficiently strong, so that the cyclotron radius $r_c\simeq v_F/\omega_c$ is much smaller than the sample width $W$ ($v_F$ is the Fermi velocity of electrons), electrons in the bulk of the 2DEG rotate around the cyclotron orbits. If the center of the cyclotron orbit is close to the edge of the sample, then, hitting the edge, the electrons move very quickly in the direction from the source to the drain on one edge and in the opposite direction on the other edge. The speed of electrons on these skipping orbits is on the order of Fermi velocity. Quantum mechanically, these skipping electron orbits correspond to the Landau levels bent upward near the 2DEG edge. The speed of the electrons at higher bent up Landau levels is higher than at lower ones (see Figure \ref{fig:energy&eigenfunc}(b) below).

Now, if we put two antenna wings on the Hall bar in such a way that the incident radiation is focused only on one edge of the 2DEG, as shown in Fig. \ref{fig:geom}(b), the electrons on the irradiated edge of the sample will absorb photons and get an additional velocity on the way from the source to the drain. The absolute value of the velocity change is about Fermi velocity. The electrons on the opposite side of the 2DEG are not irradiated and their velocity remains unchanged. As a result, a strong photocurrent is generated in the proposed detector geometry.

We emphasize that in the design proposed here, the problem of creating asymmetry in the illuminated sample is easily solved. As is known, if a symmetric sample is irradiated by normally incident radiation, the generated photocurrent or photovoltage are always equal to zero, since the currents generated in the forward and reverse directions are the same. Therefore, in Ref. \cite{Dyakonov96}, for example, different boundary conditions were imposed on the source and drain contacts to the 2DEG. In Ref. \cite{Giliberti15} a narrow gate had to be  located asymmetrically relative to the center of the 2DEG sample. In Ref. \cite{Popov2011} a complicated  double-grating-gate structure with an asymmetric unit cell was proposed. In all these cases, the currents generated in the forward and reverse directions are different due to asymmetry, and the resulting photocurrent is obtained as a small difference of the two large currents. In the case proposed in this paper, the currents flowing from the source to the drain and in the opposite direction are spatially separated by a macroscopic distance (the sample width $W$) due to the magnetic field, and the radiation can be precisely focused on only one flow. Therefore, the proposed technique should potentially give a significantly stronger response compared to the case of zero magnetic field.

Now let us present our theory (Section \ref{sec:theory}) and its results (Section \ref{sec:results}). Section \ref{sec:conclud} contains some discussions and conclusions.

\section{Theory\label{sec:theory}}

\subsection{Single-particle spectrum and eigenfunctions of 2D electrons\label{sec:spectrum}}

In our theory, we will not take into account the effects of Coulomb interaction which can lead to the appearance of alternating stripes of compressible and incompressible electron liquids near the edge of the sample \cite{Chklovskii92}. This more complex problem can be considered later if the experiment shows a discrepancy between the measured data and the predictions of the present single-particle theory.

With the aim of describing the behavior of the Hall-bar sample only at the left (irradiated) edge, Figure \ref{fig:geom}(b), we consider a 2DEG occupying a half-plane $z=0$, $x>0$, in a perpendicular magnetic field $\bm B=(0,0,B)$. The motion of the 2D electrons is described by Schr\"odinger equation
\be
\hat H_0\psi({\bm r})=\frac{1}{2m}\left(\hat {\bm p}+\frac {|e|}{c}\bm A\right)^2\psi({\bm r})=E\psi({\bm r})
\label{SchrEqPsi}
\ee 
with the boundary conditions $\psi(0,y)=0$ and $\psi(+\infty,y)=0$. For the vector potential we use the Landau gauge $\bm A=(0,Bx,0)$ and search for a solution in the form
\be 
\psi(x,y)=e^{iky}\phi(x),\label{psi-phi}
\ee
where $k\equiv k_y$ is the electron wave vector in the $y$-direction. The function $\phi(x)$ satisfies the equation
\be
-\frac{\hbar^2}{2m} \frac{\p^2 \phi(x)}{\p x^2}+\frac{m\omega_c^2}{2}\left(x+l^2k\right)^2\phi(x)=E\phi(x),
\label{SchrEqPhi}
\ee 
where $l=\sqrt{\hbar c/|e|B}=\sqrt{\hbar/m\omega_c}$ is the magnetic length and $X_k=-l^2k$ the oscillator center. Introducing a dimensionless parameter $z=\sqrt{2}\left(x/l+kl\right)$ and the function $\Phi(z)\equiv\phi\Big(\sqrt{2}(x+l^2k)/l\Big)$ we get the following differential equation
\be 
\Phi''(z)-\left(a+\frac{z^2}4\right)\Phi(z)=0\label{de}
\ee
for the function $\Phi(z)$. Here, the parameters $a\equiv -(\eta+ 1/2)$ and $\eta$ are related to the eigenenergy $E$ as follows:
\be 
\frac{E}{\hbar\omega_c}=\eta+\frac 12=-a.
\ee

\begin{figure}[ht!]
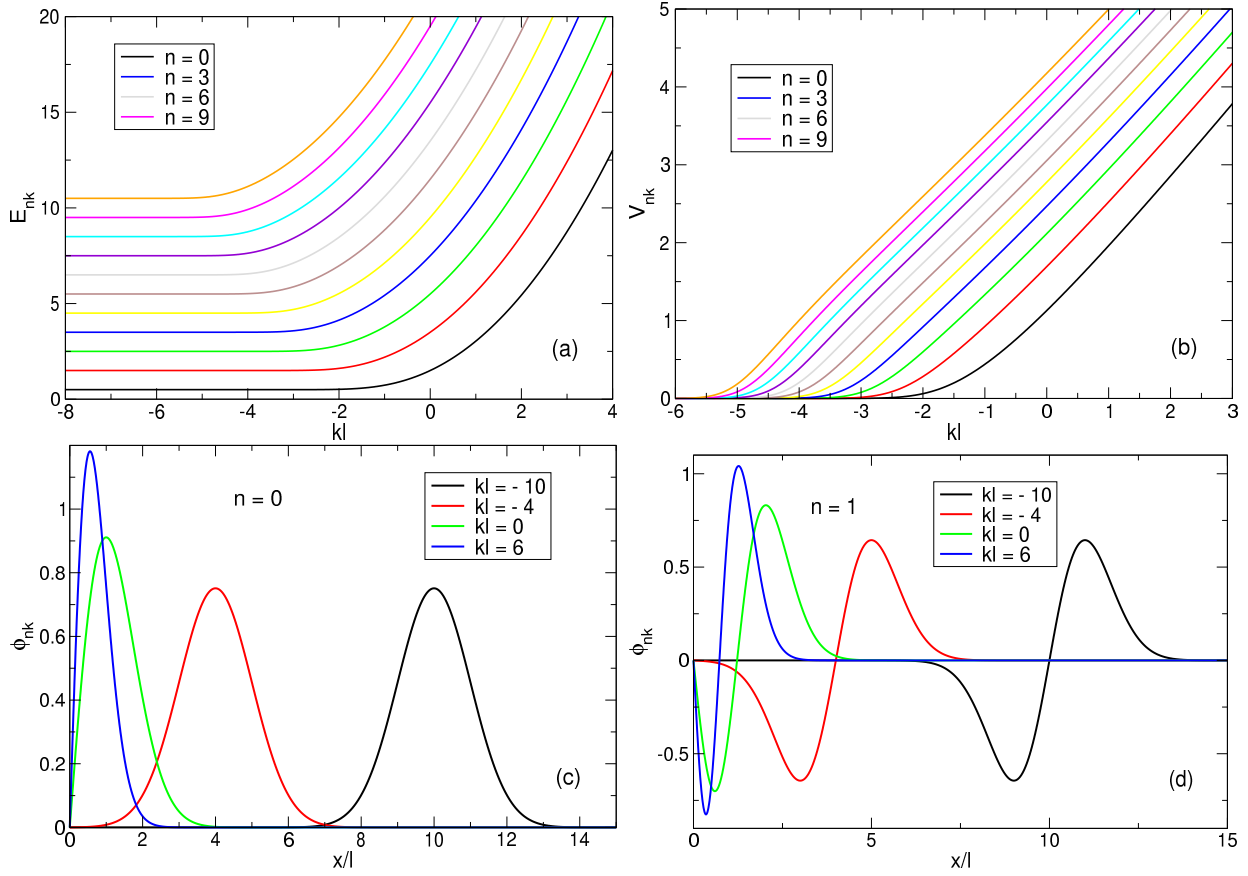

\includegraphics[width=0.49\textwidth]{fig2a.eps}
\includegraphics[width=0.49\textwidth]{fig2b.eps}
\includegraphics[width=0.49\textwidth]{fig2c.eps}
\includegraphics[width=0.49\textwidth]{fig2d.eps}
\caption{\label{fig:energy&eigenfunc}(a) The spectrum of electrons $E_{nk}$, in units $\hbar\omega_c$, and (b) their velocity $(dE_{nk}/dk)/\hbar$, in units $l\omega_c$, as a function of $kl$ near the edge of the 2D electron gas in a quantized magnetic field, for $n$ from $n=0$ to $n=10$. (c)-(d) The eigenfunctions of electrons $\phi_{nk}(x/l)$, for (c) $n=0$, (d) $n=1$, and for a few values of $kl$.}
\end{figure}

Equation (\ref{de}) is the differential equation for parabolic cylinder functions $U(a,z)$ and $V(a,z)$, see Appendix \ref{app:parcylfunc}. The general solution of Eq. (\ref{SchrEqPhi}) can be written as 
\be 
\phi(x)=C_1 U\left(-\eta-\frac 12,\sqrt{2}\left(\frac xl+kl\right)\right)+C_2 V\left(-\eta-\frac 12,\sqrt{2}\left(\frac xl+kl\right)\right),
\ee
where $C_1$ and $C_2$ are arbitrary constants. The function $V(a,x)$ tends to infinity at large positive $x$, therefore using the boundary condition at $x\to +\infty$, $\phi_{x\to +\infty}=0$, we obtain $C_2=0$. Imposing the second boundary condition at $x=0$, $\phi_{x=0}=0$, we get the eigenvalue equation
\be 
U\left(-\eta-\frac 12,\sqrt{2} kl\right)=0,
\ee
which relates the parameter $\eta$ with the normalized wavevector $kl$. Its numerical solution yields an infinite number of curves $\eta_n(kl)$, $n=0,1,\dots$, so that the spectrum of 2D electrons near the edge of the 2DEG in a quantized magnetic field assumes the form
\be 
E_{nk}=\hbar\omega_c\left(\eta_n(kl)+\frac 12\right),\ \ n=0,1,\dots.
\ee
The dependence of energy $E_{nk}$ on the Landau level index $n$ and on the normalized wave vector $kl$ is shown in Figure \ref{fig:energy&eigenfunc}(a) for $n=0,\dots,10$. The corresponding velocities 
\be 
V_{nk}=\frac 1\hbar\frac{\p E_{nk}}{\p k}=\omega_c l \frac{\p \eta_n(kl)}{\p(kl)},\ \ n=0,1,\dots,
\ee
are shown in Figure \ref{fig:energy&eigenfunc}(b). The eigenfunctions are
\be 
\phi_{nk}(x)=C_{nk}U\left(-\eta_n(kl)-\frac 12,\sqrt{2}\left(\frac xl+kl\right)\right),
\label{eigenfun-unnorm}
\ee
where $C_{nk}$ are the normalization coefficients. Figures \ref{fig:energy&eigenfunc}(c,d) show the functions $\phi_{nk}(x)$ for $n=0$ and $n=1$ at a few values of the quantum number $k$. One sees that, if $kl$ is negative and large in absolute value, $|kl|\gg 1$, i.e., the oscillator center $X_k=-l^2k$ lies inside the bulk of the 2DEG, far from the edge, the functions $\eta_n(kl)$ take non-negative integer values, $\eta_n(kl)=n\ge 0$, and the energy spectrum reproduces the usual Landau quantization, $E_{nk}=\hbar\omega_c(n+1/2)$. The eigenfunctions correspond to the conventional Landau functions in this case (black and red curves in Figs. \ref{fig:energy&eigenfunc}(c,d)). As $kl$ approaches zero and becomes positive, the Landau levels bend upward and the wave functions are localized near the 2DEG edge (green and blue curves).

The solution of Schr\"odinger equation (\ref{SchrEqPsi}) then looks as follows
\be 
|\lambda\rangle\equiv|nk\rangle\equiv\psi_{nk}(x,y)=\frac{1}{\sqrt{L_y}}e^{iky}\phi_{nk}(x),
\label{fullWF}
\ee
where $L_y$ is the length of the 2DEG in the $y$-direction. The functions $|\lambda\rangle=|nk\rangle$ are orthogonal and normalized, 
\be 
\langle \lambda|\lambda'\rangle=\langle nk|n'k'\rangle=\delta_{nn'}\delta_{kk'}.
\ee

Using the eigenstates $|\lambda\rangle$, one can calculate different matrix elements. In particular, we will need the matrix elements of the coordinate $x$ and of the velocity operator
\be 
\hat v_y=
\frac{1}{m}
\left(-i\hbar \frac{\p }{\p y}+\frac {|e|}{c} Bx\right).\label{vel-oper}
\ee
We obtain for these quantities
\be 
\langle \lambda |x|\lambda'\rangle = l\delta_{kk'}{\cal X}_{nn'}(kl),\label{MEx}
\ee
\be 
\langle \lambda |\hat v_y|\lambda'\rangle = 
\frac{\hbar}{ml}\delta_{kk'}{\cal X}_{nn'}(kl) ,\label{MEv}
\ee
where
\be 
{\cal X}_{nn'}(kl)=\int_0^{\infty} \phi_{nk}(x)\frac xl\phi_{n'k}(x)dx \label{calX}
\ee
is a dimensionless function which can be numerically calculated for any $n$, $n'$ and $kl$.

\subsection{Photocurrent response\label{sec:Phcurr}}

Now we consider our system under the action of the external time-dependent electric field, produced by the incident electromagnetic wave with the frequency $\omega$ and localized near the edge of the 2DEG. The Hamiltonian $\hat H$ assumes the form
\be 
\hat H=\hat H_0+\hat H_1,
\ee
where
\be 
\hat H_1=V(x) \cos\omega t=-e\Phi_{\rm ac}(x) \cos\omega t
\label{H1}
\ee
and the electric potential $\Phi_{\rm ac}(x)$ describes the ac electric field $E_{\rm ac}(x)=-\p \Phi_{\rm ac}(x)/\p x$ acting on electrons in the gap between the two antenna wings. The function $E_{\rm ac}(x)$ is localized at the length $\sim g$, see Figure \ref{fig:geom}(c). 

The electrodynamic response of the edge 2D electrons to the perturbation $\hat H_1$ can be described by Liouville equation for the density matrix
\be 
\frac{\p\hat\rho}{\p t}=\frac 1{i\hbar}[\hat H_0+\hat H_1,\hat\rho]-\frac{\hat\rho-\hat\rho_0}{\tau_{\rm edge}},
\ee
where we have introduced a phenomenological time $\tau_{\rm edge}$ which describes the relaxation of the photo-excited edge electrons to the equilibrium distribution after the excitation is switched off. This relaxation time is not identical to the conventional momentum relaxation time $\tau_{p}$ which enters Drude formulas for the bulk conductivity and determines, for example, the linewidth of the 2D bulk plasmons and magnetoplasmons. Since skipping along the edge 2D electrons have no opportunity to scatter back and can therefore run without scattering over macroscopically large distances, this time is much larger than $\tau_{p}$. We believe that the value of $\omega_c\tau_{\rm edge}$ can be estimated from the ratio $\rho_{xy}/\rho_{xx}$, where $\rho_{xy}$ and $\rho_{xx}$ are the experimentally measured Hall and longitudinal resistivities. Therefore, $\tau_{\rm edge}/\tau_{p}$ is significantly larger than 1 on the quantum Hall plateaus and approaches 1 when Landau quantization is not seen in transport measurements (e.g., at elevated temperatures). We will consider $\tau_{\rm edge}$ as an unknown phenomenological parameter that should be estimated by comparing theoretical results with experiments. 

The unperturbed Hamiltonian and the equilibrium density matrix satisfy the eigenvalue equations
\be 
\hat H_0|\lambda\rangle=E_\lambda|\lambda\rangle,\ \ \ \hat\rho_0|\lambda\rangle=f_\lambda|\lambda\rangle ,
\ee
where 
\be 
f_\lambda=\frac 1{1+\exp\left(\frac{E_\lambda-\mu}T\right)}
\ee
is the equilibrium Fermi distribution function, $\mu$ is the chemical potential and $T$ is the temperature. Substituting $\hat\rho=\hat\rho_0+\hat\rho_1+\hat\rho_2+\dots$, where $\hat\rho_n$ is proportional to the $n$-th power of the ac electric field, we get in the first and second orders
\be 
i\hbar\left(\frac{\p\hat\rho_1}{\p t}+\frac{\hat\rho_1}{\tau_{\rm edge}}\right)=[\hat H_0,\hat\rho_1]+[\hat H_1,\hat\rho_0],
\ee
\be 
i\hbar\left(\frac{\p\hat\rho_2}{\p t}+\frac{\hat\rho_2}{\tau_{\rm edge}}\right)=[\hat H_0,\hat\rho_2]+[\hat H_1,\hat\rho_1].
\ee
The first order solution for the density matrix has the form
\be 
\langle\lambda |\hat\rho_1|\lambda'\rangle  =\frac {1}2 \langle\lambda |V(x)|\lambda'\rangle 
\left(
\frac{f_{\lambda}-f_{\lambda'}}{E_\lambda-E_{\lambda'}-\hbar\omega -i\hbar \gamma_{\rm edge}}e^{-i\omega t} +
\frac{f_{\lambda}-f_{\lambda'}}{E_\lambda-E_{\lambda'}+\hbar\omega -i\hbar \gamma_{\rm edge}}e^{i\omega t}
\right),
\ee
where $\gamma_{\rm edge}=1/\tau_{\rm edge}$ and $\langle\lambda |V(x)|\lambda'\rangle$ are the matrix elements of the potential energy $V(x)$ with the wave functions (\ref{fullWF}). In the second order, we get the contributions to $\langle\lambda |\hat\rho_2|\lambda'\rangle$, oscillating with the frequency $2\omega$ and time-independent terms. Since we are only interested in the dc photocurrent response, we calculate only the latter contribution and obtain the following result for the second-order density matrix:
\ba 
\langle \lambda |\hat\rho_2^{(0)}|\lambda'\rangle  &=&\frac 14
\frac{1}{E_{\lambda'}-E_{\lambda}+i\hbar \gamma_{\rm edge}} 
\sum_{\lambda''}\langle\lambda |V(x) |\lambda''\rangle \langle\lambda'' | V(x)|\lambda'\rangle 
\nonumber \\ &\times&
\Bigg[ 
\Bigg( \frac{f_{\lambda''}-f_{\lambda'}}{E_{\lambda''}-E_{\lambda'}-\hbar\omega -i\hbar \gamma_{\rm edge}} +\frac{f_{\lambda''}-f_{\lambda'}}{E_{\lambda''}-E_{\lambda'}+\hbar\omega -i\hbar \gamma_{\rm edge}}
\Bigg)
\nonumber \\ &-&
\Bigg( \frac{f_{\lambda}-f_{\lambda''}}{E_{\lambda}-E_{\lambda''}-\hbar\omega -i\hbar \gamma_{\rm edge}}
+\frac{f_{\lambda}-f_{\lambda''}}{E_{\lambda}-E_{\lambda''}+\hbar\omega -i\hbar \gamma_{\rm edge}}\Bigg)
 \Bigg];
\label{rho2ndOrder}
\ea
the superscript $^{(0)}$ indicates that this is only the time-independent contribution to the density matrix.

Now we need to calculate the second-order photocurrent generated by the localized ac electric field $E_{\rm ac}(x)$ in the gap between the gates. Using the definition of the current density
\be 
j_y(\bm r_0)= -\frac e2 {\rm Sp} \Big[\left(\hat v_y\delta(\bm r-\bm r_0)+\delta(\bm r-\bm r_0)\hat v_y\right)\hat\rho \Big],
\ee
where $\hat v_y$ is the velocity operator (\ref{vel-oper}), and integrating the result over $dx$ from $x=0$ up to $x=\infty$, we finally get the following expression for the total photocurrent in the $y$-direction
\ba 
I_2^{(0)}&=&
-\frac {eg_s\omega_cl}{4L_y}\sum_{k}   \sum_{nn'}  
\frac{{\cal X}_{nn'}(kl)}{E_{nk}-E_{n'k}+i\hbar \gamma_{\rm edge}} 
\sum_{n''}  {\cal V}_{n'n''}(kl) {\cal V}_{n''n}(kl) 
\nonumber \\ &\times&
\Bigg[ 
\Bigg( \frac{f_{n''k}-f_{nk}}{E_{n''k}-E_{nk}-\hbar\omega -i\hbar \gamma_{\rm edge}} +\frac{f_{n''k}-f_{nk}}{E_{n''k}-E_{nk}+\hbar\omega -i\hbar \gamma_{\rm edge}}
\Bigg)
\nonumber \\ &+&
\Bigg( \frac{f_{n''k}-f_{n'k}}{E_{n'k}-E_{n''k}-\hbar\omega -i\hbar \gamma_{\rm edge}}
+\frac{f_{n''k}-f_{n'k}}{E_{n'k}-E_{n''k}+\hbar\omega -i\hbar \gamma_{\rm edge}}\Bigg)
 \Bigg],
\label{2ndOrderCurrent}
\ea
where $g_s=2$ is the spin degeneracy factor.

Equation (\ref{2ndOrderCurrent}) is the central result of our work. It gives the second-order photocurrent generated by the incident electromagnetic wave focused at the 2DEG edge in the gap between two antenna wings, Figs. \ref{fig:geom}(b,c). Apart from the energy levels $E_{nk}$ and the matrix elements of the coordinate ${\cal X}_{nn'}$, the photocurrent also depends on the matrix elements 
\be 
\langle \lambda |V(x)|\lambda'\rangle = \delta_{kk'}{\cal V}_{nn'}(kl),\label{MEV}
\ee
\be 
{\cal V}_{nn'}(kl)=\int_0^{\infty} \phi_{nk}(x)V(x)\phi_{n'k}(x)dx,
\label{calV}
\ee 
of the ac potential of the incident electromagnetic wave $V(x)$. To calculate these matrix elements we need a specific model of the ac potential $V(x)$ in the gap between the antenna wings.

\subsection{Model of the ac potential $V(x)$\label{sec:Vmodel}} 

\begin{figure}[ht!]
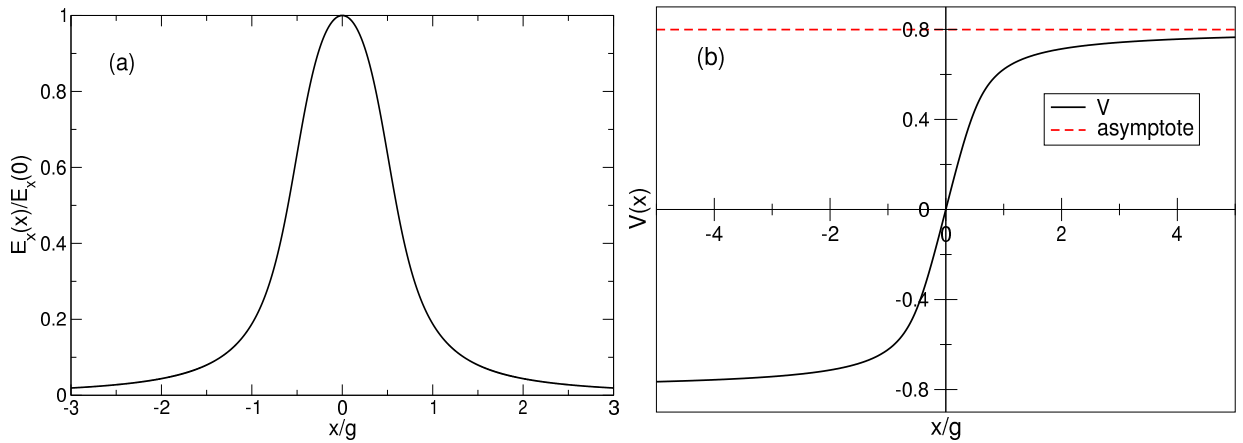

\includegraphics[width=0.49\textwidth]{fig3a.eps}
\includegraphics[width=0.49\textwidth]{fig3b.eps}
\caption{\label{fig:field} (a) The model ac electric field (\ref{Exmodel}) and (b) the corresponding potential energy (\ref{Vmodel}) in the gap between two gates, near the edge of the 2DEG; $G=g/d=3$.}
\end{figure}

We assume that the two antenna wings are located near the edge of the 2DEG as shown in Figures \ref{fig:geom}(b,c). The distance between the antenna wings is $g$, the distance between the 2DEG and the right gate is $d$, and the 2DEG edge is located just in the middle between the gates. For typical experimental parameters $d\lesssim 100$ nm and $g\lesssim 400$ nm, the distribution in the ac electric field in the near-gap region was studied using numerical simulations in Ref. \cite{Chen24}. The numerical results for the ac electric field $E_x(x)$ in the plane of the 2D gas, obtained in that paper, can be well described by the following analytical formula
\be 
E_x\left(x,b,d\right)= E_x\left(0,b,d\right)
\frac {\arctan\left(\frac{g/2+x}{d}\right)+ \arctan\left(\frac{g/2-x}{d}\right)}
{2\arctan\left(\frac{g}{2d}\right)},\label{Exmodel}
\ee
see Figure \ref{fig:field}(a). The potential energy $V(x,b,d)=-e\Phi(x,b,d)$ can then be obtained by integrating over $dx$, 
\be 
V(x,b,d)=e\int_0^x E_x\left(x',b,d\right)dx' =
eE_{\rm ac}g{\cal G}\left(X,G\right),
\label{Vmodel}
\ee
where $E_{\rm ac}$ is the ac electric field in the center of the gap at the edge of the 2DEG, $X=x/g$, $G=g/d$, and the function ${\cal G}\left(X,G\right)$ is defined as follows
\be 
{\cal G}\left(X,G\right)=\frac{F\left[G/2+GX\right]-F\left[G/2-GX\right]}{2G\arctan(G/2)},
\ee
where
\be 
F(z)=z\arctan z-\frac 12\ln\left(1+z^2\right),
\ee
see Figure \ref{fig:field}(b). To calculate the matrix elements ${\cal V}_{nn'}$ in Eq. (\ref{calV}) we use the model (\ref{Exmodel})-(\ref{Vmodel}). Apart from the dimensionless wave vector $kl$, the matrix elements ${\cal V}_{nn'}(kl,g/d,l/g)$ also depend on the two dimensionless parameters $G=g/d$ and $L=l/g$. The latter can be represented as the square root of the ratio of the two energies, $\hbar^2/mg^2$ and $\hbar\omega_c$,
\be 
L=\frac lg=\sqrt{H},
\ee
where
\be 
H=\frac {\hbar^2/mg^2}{\hbar\omega_c}=\frac {\hbar}{mg^2\omega_c}.\label{H}
\ee

\section{Results\label{sec:results}}

\subsection{General formula and choice of parameters}

Let us now present the results of our theory. First of all, we notice that the photocurrent (\ref{2ndOrderCurrent}) depends on the geometrical factor $G=g/d$ and is a function of the following six energy parameters:
\be 
\mu,\ \ \ T,\ \ \ \hbar\omega,\ \ \ \hbar\omega_c, \ \ \ \hbar\gamma_{\rm edge} ,\ \ \ \frac{\hbar^2}{mg^2}. \label{params}
\ee
To clarify the physical meaning of the results, it is convenient to normalize all energy arguments to $\hbar\omega_c$. Then the photocurrent (\ref{2ndOrderCurrent}) can be represented as follows

\be 
I_2^{(0)}=
\frac{eg_s\omega_c}{4\pi}
\left(\frac {eE_{\rm ac}g}{\hbar\omega_c}\right)^2
{\cal J}_{\rm ph,edge} \left(\frac{\omega}{\omega_c},\frac{\gamma_{\rm edge}}{\omega_c},\frac{\mu}{\hbar\omega_c},\frac{T}{\hbar\omega_c},H,\frac gd\right),
\label{2ndOrderCurrent-Ww}
\ee
The first factor here has the dimensionality of Ampere. If we assume that the experiment will be carried out in a GaAs/AlGaAs heterostructure ($m=0.067m_0$) and take $B=4$ T as a typical magnetic field, then the first factor is equal to $0.27$ $\mu$A. The second factor is the perturbation theory parameter. It should be smaller than 1. This factor depends on the power of the incident electromagnetic wave and the focusing characteristics of the antenna. To estimate this factor, we adopt the experimental parameters of Ref. \cite{Michailow22}. The value of $eE_{\rm ac}g$ was estimated there as $\sim 2.7$ meV. For the magnetic field $B=4$ T and the effective mass $m=0.067m_0$ ($\hbar\omega_c=6.89$ meV), the second factor is then $0.15$, and the product of the first and second factors is 40.5 nA.

The third factor ${\cal J}_{\rm ph,edge}$ is a dimensionless function of all the essential physical parameters on which the magnetic field-induced edge photocurrent depends. Below we study the factor ${\cal J}_{\rm ph,edge}$ as a function of frequency $\omega/\omega_c$ and the chemical potential $\mu/\hbar\omega_c$, at a number of typical values of other parameters (\ref{params}). 

When choosing the numerical values of $\gamma_{\rm edge}$, $T$, $H$, etc., see Figures \ref{fig:Iph1} and \ref{fig:Iph2} below, we assume $B=4$ T and take geometrical parameters close to those used in the IPPE experiment \cite{Michailow22}. The gap width $g$ and the distance between the gate and the 2D gas $d$ were equal $g=270$ nm and $d=90$ nm, respectively, so that we take the values close to $G=3$ for the ratio $G=g/d$. The parameter (\ref{H}) at $B=4$ T equals $H=0.00225$ for $g=270$ nm and $H=0.009$ for a two times smaller $g$. The temperature parameters $T/\hbar\omega_c=0.05$ and 1 correspond to $T=4$ K and $T=80$ K, respectively. The cyclotron resonance $\omega=\omega_c$ at $B=4$ T corresponds to the frequency $f=1.67$ THz, therefore the maxima of the photocurrent at all pictures shown below are seen approximately at $f=2$ THz, the frequency used in Ref. \cite{Michailow22}. 

It is important to choose a relevant value for the parameter $\gamma_{\rm edge}/\omega_c$ because the absolute value of the photocurrent very strongly depends on $\gamma_{\rm edge}$. If $B=4$ T the product $\omega_c\tau_p$ in samples with good electron mobility can easily reach values $\omega_c\tau_p\sim 10$. As we discussed above, the time $\tau_{\rm edge}$ can be substantially longer than $\tau_p$. At the Hall plateaus and low temperatures the ratio $\tau_{\rm edge}/\tau_p$ can be as large as $\sim 100$ or larger, then the parameter $\gamma_{\rm edge}/\omega_c$ can be as low as $10^{-3}$. For our plots, we take more conservative values around $\gamma_{\rm edge}/\omega_c= 0.01$.

\begin{figure}[ht!]
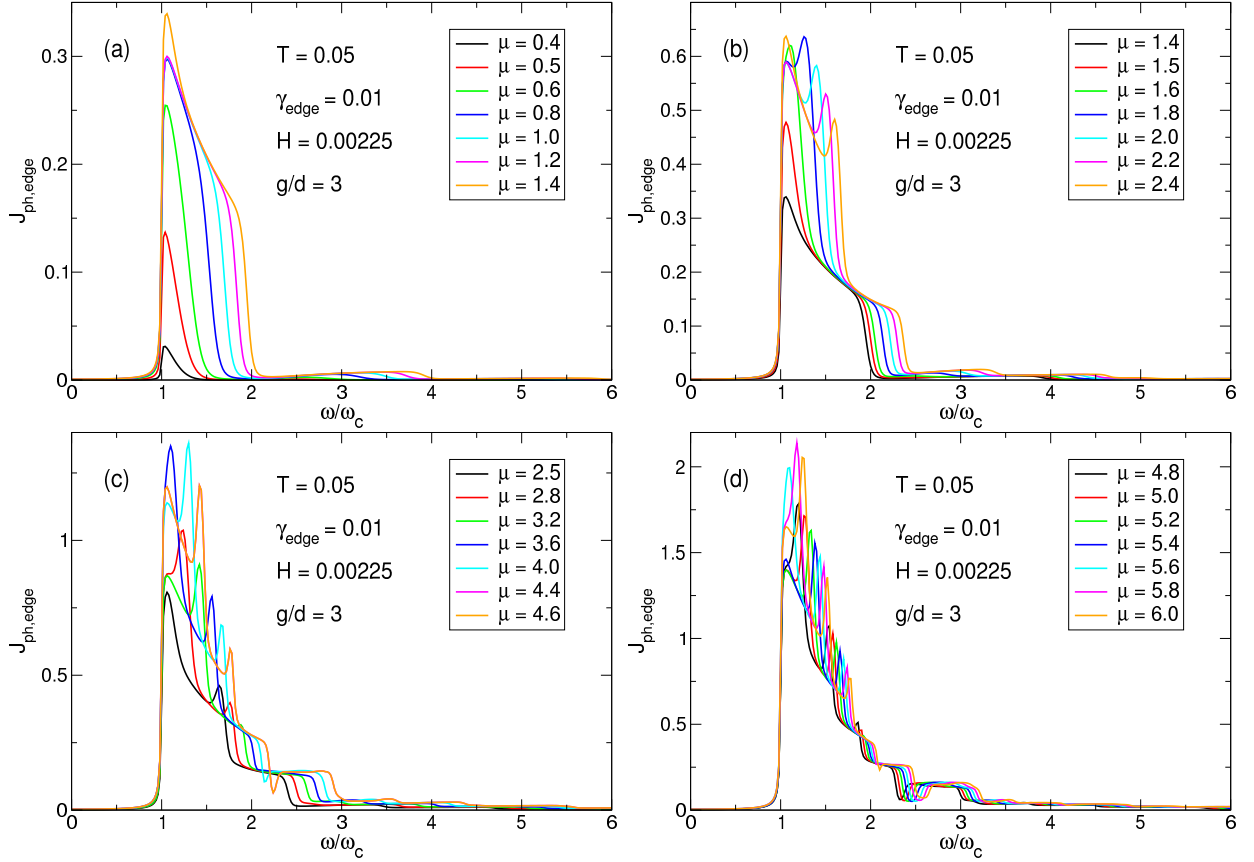

\includegraphics[width=0.49\textwidth]{fig4a.eps}
\includegraphics[width=0.49\textwidth]{fig4b.eps}
\includegraphics[width=0.49\textwidth]{fig4c.eps}
\includegraphics[width=0.49\textwidth]{fig4d.eps}
\caption{\label{fig:Iph1} (a)--(d) The photocurrent factor ${\cal J}_{\rm ph,edge}$ defined in Eq. (\ref{2ndOrderCurrent-Ww}) as a function of frequency $\omega/\omega_c$ at different values of the chemical potential $\mu/\hbar\omega_c$ varying from 0.4 up to 6. Other parameters are $\gamma_{\rm edge}/\omega_c=0.01$, $T/\hbar\omega_c=0.05$, $H=0.00225$, and $g/d=3$.}
\end{figure}

\subsection{Frequency spectra of photocurrent at different electron densities}

Let us now consider the frequency dependence of the factor ${\cal J}_{\rm ph,edge}$ at different values of the chemical potential $\mu /\hbar\omega_c$. Figure \ref{fig:Iph1}(a) shows this function at small values of $\mu$ and at the parameters shown in the Figure caption. If the lowest Landau level is only slightly occupied, so that $\mu/\hbar\omega_c=0.4$ and $f_{0k}\approx 0.12$ in the bulk of the 2DEG, then it has the form of an asymmetric resonance with a maximum at $\omega$ slightly greater than $\omega_c$. The height of the resonance is rather small ($\simeq 0.03$). As the chemical potential increases, the resonance becomes wider and higher. If $\mu/\hbar\omega_c = 0.5$, so that the lowest Landau level is half occupied, $f_{0k}\approx 0.5$, the resonance height increases more than fourfold (red curve) and becomes equal to $\sim 0.137$. When the chemical potential becomes equal to $\mu/\hbar\omega_c = 1.4$, so that $f_{0k}\approx 1$ and $f_{1k}\approx 0.12$, the height of the photocurrent resonance becomes already equal to $\sim 0.34$. 

\begin{figure}[ht!]
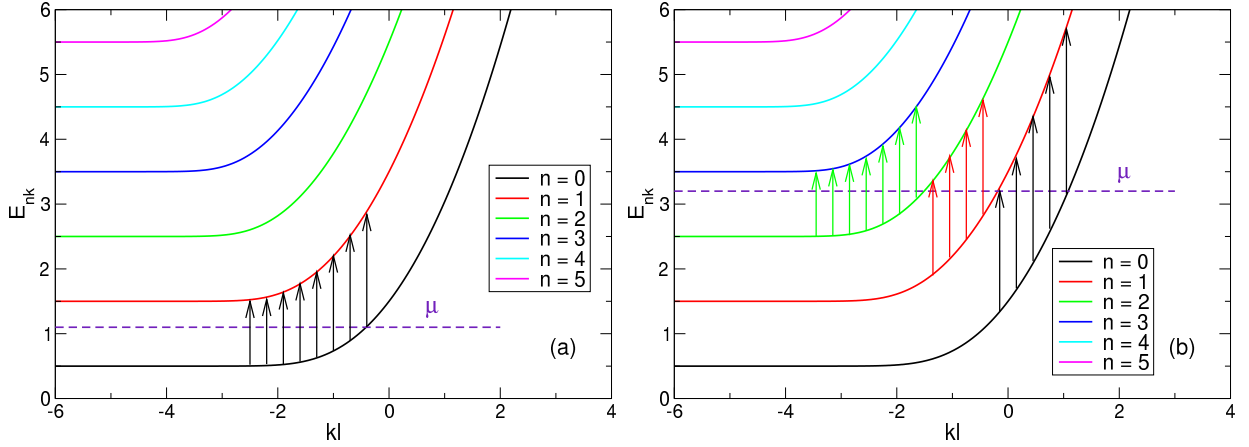

\includegraphics[width=0.49\textwidth]{fig5a.eps}
\includegraphics[width=0.49\textwidth]{fig5b.eps}
\caption{\label{fig:edgetrans} The vertical transitions between the nearest Landau levels $n\to n+1$ for (a) $\mu/\hbar\omega_c=1.1$ and (b) $\mu/\hbar\omega_c=3.2$. The transitions $n\to n'$ with $n'-n>1$ are allowed near the edge but not shown in the Figure.}
\end{figure}

The frequency range in which photocurrent is efficiently generated is significantly wider than that of conventional cyclotron resonance and lies between $\omega\sim\omega_c$ and $\omega\sim 2\omega_c$, Fig. \ref{fig:Iph1}(a). This is due to transitions between the edge states of the zeroth and first Landau levels, which are possible for edge electrons in a wider range than in the bulk, Fig. \ref{fig:edgetrans}(a). Figure \ref{fig:Iph1}(a) also shows a weak photocurrent response at frequencies up to $\omega\sim 4\omega_c$. This is due to transitions between Landau levels $n\to n'$ with $n'-n>1$. Such transitions are forbidden in the bulk, but are possible near the edge.

Figure \ref{fig:Iph1}(b) shows the function ${\cal J}_{\rm ph,edge}$ for larger values of the chemical potential $\mu$ from $\mu/\hbar\omega_c= 1.4$ to 2.4. One sees that when Landau level with $n = 1$ is filled in the 2DEG bulk, an additional resonance appears in the photocurrent spectrum, see the blue to orange curves corresponding to $\mu/\hbar\omega_c= 1.8$, 2.0, 2.2 and 2.4. As the chemical potential increases further, Figs. \ref{fig:Iph1}(c) and (d), the structure of the photocurrent spectrum becomes more complex, with a large number of narrow resonances appearing each time a new Landau level is filled. For example, the cyan curve in Figure \ref{fig:Iph1}(c) corresponding to $\mu/\hbar\omega_c= 4.0$ has three maxima at $\omega/\omega_c= 1.05$, 1.28 and 1.66 and a plateau-like behavior between $\omega/\omega_c=2.19$ and 2.74, as well as at $\omega/\omega_c$ bigger than 2.86. The orange curve in Figure \ref{fig:Iph1}(d) corresponding to $\mu/\hbar\omega_c=6.0$ has local maxima at $\omega/\omega_c= 1.05$, 1.24, 1.51, 1.76, and a plateau-like behavior in the range from $\omega/\omega_c=2.15$ up 2.43 and from $\omega/\omega_c=2.71$ up to 3.2. The possible transitions between Landau levels $n$ and $n+ 1$ are shown in Fig. \ref{fig:edgetrans}(b) for $\mu/\hbar\omega_c= 3.2$. The transitions $n\to n'$ with $n'-n>1$ are allowed near the edge too but are not shown in Figure \ref{fig:edgetrans}.

\begin{figure}[ht!]
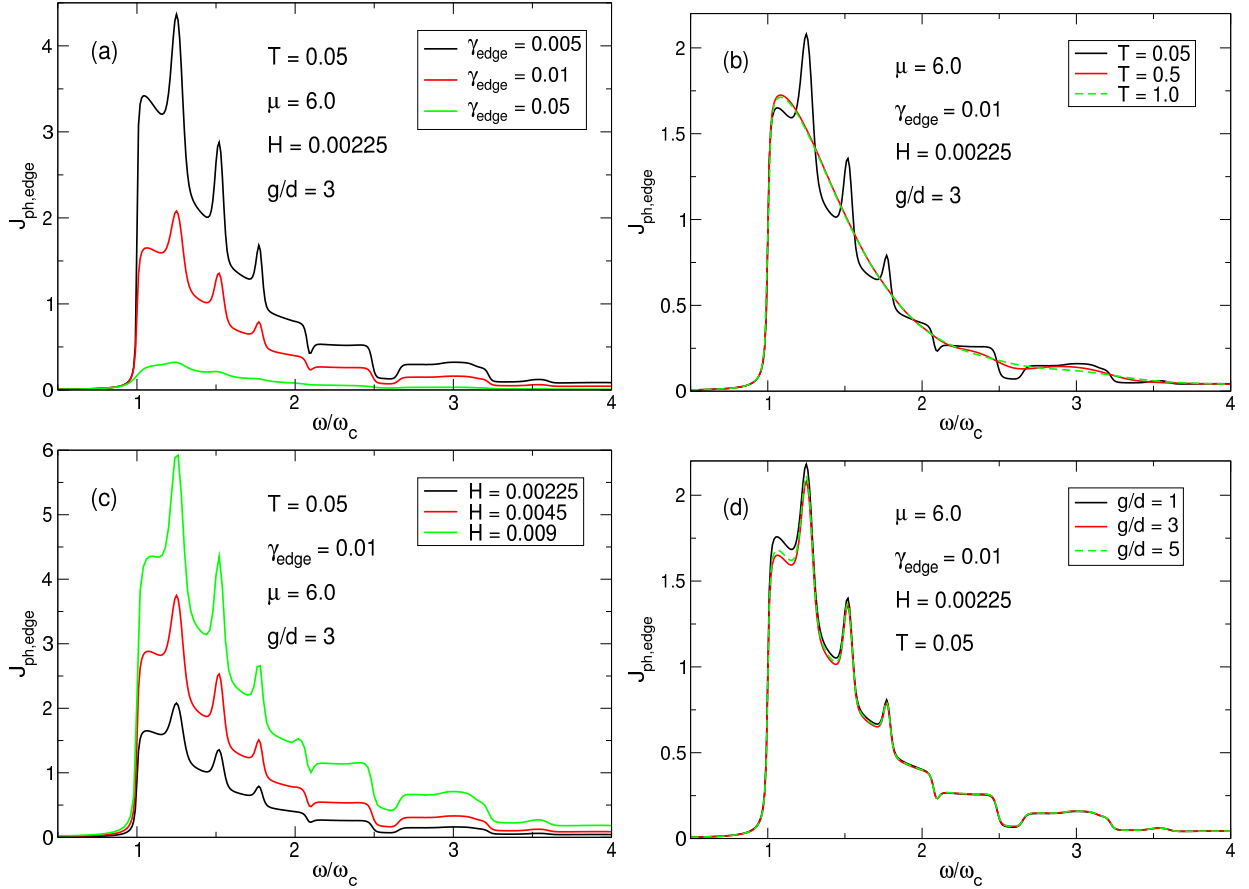

\includegraphics[width=0.49\textwidth]{fig6a.eps}
\includegraphics[width=0.49\textwidth]{fig6b.eps}
\includegraphics[width=0.49\textwidth]{fig6c.eps}
\includegraphics[width=0.49\textwidth]{fig6d.eps}
\caption{\label{fig:Iph2} Influence of different input parameters on the photocurrent factor ${\cal J}_{\rm ph,edge}$: (a) different values of $\gamma_{\rm edge}/\omega_c$, (b) different temperatures, (c) different parameters $H$, and (d) different values of $g/d$.}
\end{figure}

\subsection{Dependencies of the photocurrent on the scattering rate, temperature and geometric parameters of the structure}

Let us now consider how the parameters $\gamma_{\rm edge}/\omega_c$, $T/\hbar\omega_c$, $H$, and $g/d$ influence the photocurrent spectra. Figure \ref{fig:Iph2}(a) shows the function ${\cal J}_{\rm ph,edge}$ at different values of the scattering rate $\gamma_{\rm edge}/\omega_c$ and at $\mu/\hbar\omega_c=6$, $T/\hbar\omega_c=0.05$, $H=0.00225$, and $g/d=3$. The overall shape of the photocurrent spectra remains roughly the same at different values of $\gamma_{\rm edge}$, but the height of the photoresponse function varies substantially. Roughly, the photocurrent is inversely proportional to $\gamma_{\rm edge}$, which can be also shown analytically. Since the number $\gamma_{\rm edge}/\omega_c=0.01$ that we used in Figure \ref{fig:Iph1} is in fact conservative, as we discussed above, the photoresponse could be an order of magnitude greater in system with $\gamma_{\rm edge}/\omega_c\sim 10^{-3}$.

Figure \ref{fig:Iph2}(b) illustrates the influence of temperature $T/\hbar\omega_c$ on the photocurrent factor ${\cal J}_{\rm ph,edge}$; other parameters are assumed  to remain unchanged: $\mu/\hbar\omega_c=6$, $\gamma_{\rm edge}/\omega_c=0.01$, $H=0.00225$, and $g/d=3$. Since the factor $\gamma_{\rm edge}/\omega_c$ is kept constant, this plot illustrates the influence of the Fermi function smearing on the factor ${\cal J}_{\rm ph,edge}$. When the temperature is low, $T/\hbar\omega_c=0.05$, the photocurrent spectrum consists of several narrow lines, the two highest of which are at $\omega/\omega_c=1.06$ and 1.25. The values of ${\cal J}_{\rm ph,edge}$ at these points are 1.65 and 2.08, respectively. When the temperature increases by a factor of ten, from $T/\hbar\omega_c=0.05$ to $T/\hbar\omega_c=0.5$, the narrow low-temperature resonances are smeared, but the overall shape of the photoresponse curve remains essentially the same. The maximum of the ${\cal J}_{\rm ph,edge}$ spectrum is now at $\omega/\omega_c=1.08$ and is ${\cal J}_{\rm ph,edge}\approx 1.725$. 

Remarkably, despite a ten-fold increase in temperature, the photoresponse amplitude decreases by only about 17 \%. Moreover, when the temperature increases further, from $T/\hbar\omega_c=0.5$ to $T/\hbar\omega_c=1$, the amplitude of the photoresponse practically does not change: at $T/\hbar\omega_c=1$ it is ${\cal J}_{\rm ph,edge}\approx 1.714$. Numerically, $T/\hbar\omega_c=1$ corresponds to $T\sim 80$ K, according to our estimates above. Therefore, if only the influence of the Fermi function smearing on the factor ${\cal J}_{\rm ph,edge}$ is considered, Figure \ref{fig:Iph2}(b), the function ${\cal J}_{\rm ph,edge}$ should remain approximately the same even at room temperature, when $T/\hbar\omega_c\sim 3$. 

\begin{figure}[ht!]
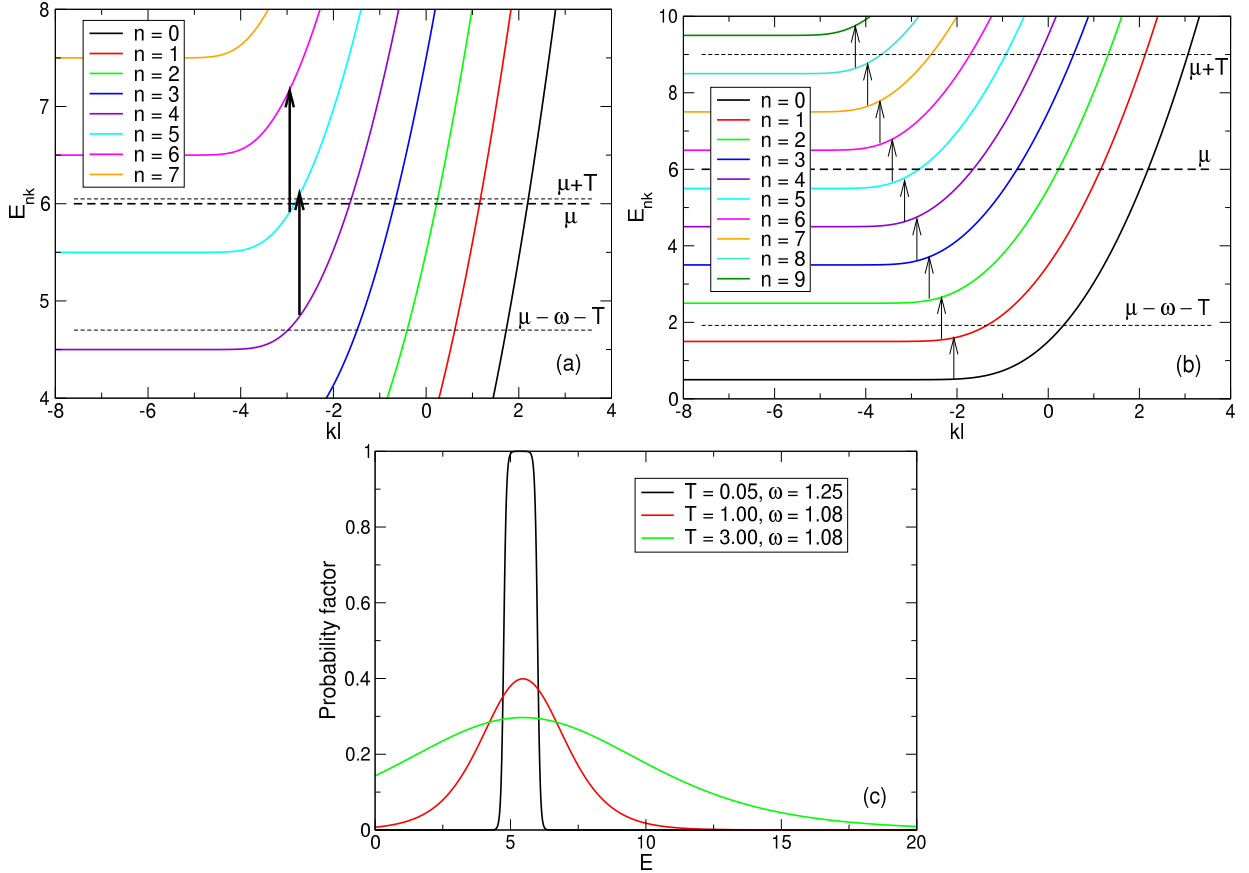

\includegraphics[width=0.49\textwidth]{fig7a.eps}
\includegraphics[width=0.49\textwidth]{fig7b.eps}
\includegraphics[width=0.49\textwidth]{fig7c.eps}
\caption{\label{fig:Tdependence} Allowed vertical transitions at (a) temperature $T/\hbar\omega_c= 0.05$ and frequency $\omega/\omega_c=1.25$, and (b) temperature $T/\hbar\omega_c= 3$ and frequency $\omega/\omega_c=1.08$. The frequencies correspond to the photocurrent maxima, see Figure \ref{fig:Iph2}(b). The thick dashed horizontal line represents the chemical potential $\mu$, thin dashed horizontal lines correspond to the levels $\mu+T$ and $\mu-\hbar\omega-T$, thicker arrows indicate a higher transition probability. (c) The probability factor ${\cal P}$, Eq. (\ref{probfact}), as a function of energy at temperatures $T/\hbar\omega_c= 0.05$, 1, and 3, and frequencies $\omega/\omega_c$ corresponding to the maxima of the function ${\cal J}_{\rm ph,edge}$. The chemical potential equals $\mu/\hbar\omega_c=6$ on all the plots.}
\end{figure}

Why does the smearing of the Fermi function with increasing temperature have no significant influence on the function ${\cal J}_{\rm ph,edge}$? Qualitatively, this can be understood as follows. The probability of vertical optical transitions from the initial energy level $E$ to the final energy level $E+\hbar\omega$ is determined by the factor
\be 
{\cal P}=f(E,T)\left[1-f(E+\hbar\omega,T)\right],\label{probfact}
\ee
which requires that the initial state is occupied and the final state is empty; here $f(E,T)$ is the Fermi function. If the temperature is low, $T\ll\mu$, Figure \ref{fig:Tdependence}(a), and the Fermi function is close to a step function, the probability ${\cal P}$ is close to 1, but the initial energies $E$ can lie only in a narrow band $\mu-\hbar\omega-T\lesssim E\lesssim \mu+T$, see Figure \ref{fig:Tdependence}(c), black curve. Then, for given $\mu/\hbar\omega_c=6$ and $\omega/\omega_c=1.25$, only two transitions, from the energy level with $n=4$ to the energy level with $n=5$, and from the energy level $n=5$ to the energy level $n=6$ are possible, see Figure \ref{fig:Tdependence}(a). If the temperature is high, Figure \ref{fig:Tdependence}(b), the probability ${\cal P}$ of individual transitions is low ($\lesssim 0.3$ for $T/\hbar\omega_c=3$ and $\omega/\omega_c=1.08$, see Figure \ref{fig:Tdependence}(c), green curve), which is illustrated by thinner arrows on Figure \ref{fig:Tdependence}(b), but this is compensated by a large number of possible transitions $n\to n+1$: the initial energy $E$ can lie in a much broader range including the areas below the energy $E=\mu-\hbar\omega-T$ and well above the energy $E=\mu+T$. 

Thus, the smearing  of the distribution function with increasing temperature does not significantly affect the function ${\cal J}_{\rm ph,edge}$. Of greater importance in a real physical system will be the change in the function ${\cal J}_{\rm ph,edge}$ due to the increase in the parameter $\gamma_{\rm edge}/\omega_c$ with increasing temperature. While at low temperatures, especially in the quantum Hall effect regime, one can expect the parameter $\gamma_{\rm edge}/\omega_c$ to be of the order of 0.001 or lower, at room temperature, $\tau_{\rm edge}$ will be close to $\tau_p$, and hence the parameter $\gamma_{\rm edge}/\omega_c$ will increase significantly. As seen from Figure \ref{fig:Iph2}(a), the function ${\cal J}_{\rm ph,edge}$ is approximately inversely proportional to $\gamma_{\rm edge}/\omega_c$, therefore it is this parameter that will determine the temperature dependence of the photocurrent at elevated temperatures. 

Now let us return back to Figure \ref{fig:Iph2} and consider the influence of two remaining parameters on the photocurrent factor ${\cal J}_{\rm ph,edge}$. 
Figure \ref{fig:Iph2}(c) illustrates the influence of the parameter $H$ on this function. The black curve ($H=0.00225$) corresponds to the gap width $g=270$ nm at $B=4$ T. The green curve corresponds to a two times smaller gap, $g=135$ nm. Reducing the gap size between the gates can thus lead to a significant increase in the photocurrent. Note that Figure \ref{fig:Iph2}(c)  assumes that the ratio $g/d$ remains the same.

Finally, the last panel (d) of Figure \ref{fig:Iph2} shows how the photocurrent spectra change when the ratio $g/d$ varies. One sees that the parameter $g/d$ does not have a significant effect on the photoresponse.

\subsection{Estimates of the numerical value of the photocurrent}

Let us evaluate the expected absolute value of the photocurrent that can be generated in semiconductor 2D electron systems in the phenomenon discussed. As seen from the above analysis, for moderate parameters used in Figure \ref{fig:Iph1}, $H=0.00225$ and $\gamma_{\rm edge}/\omega_c=0.01$, the third factor ${\cal J}_{\rm ph,edge}$ is about 2. Together with the first and second factors in Eq. (\ref{2ndOrderCurrent-Ww}), estimated above, the expected photocurrent is about $I_2^{(0)}\approx 80$ nA which is comparable with the value ($\sim 0.14$ $\mu$A) experimentally measured in the IPPE effect at $B=0$ \cite{Michailow22}. If the gap width $g$ and the scattering rate $\gamma_{\rm edge}$ could be reduced, so that, for example, $H=0.009$ and $\gamma_{\rm edge}/\omega_c=10^{-3}$, the third factor ${\cal J}_{\rm ph,edge}$ could reach a value of $\sim 60$. Then the expected photocurrent, $\sim 2.4$ $\mu$A, could be even more than one order of magnitude greater than in \cite{Michailow22}.

So far we have normalized all energy parameters to $\hbar\omega_c$ and used for the typical magnetic field the value of $B=4$ T. The photocurrent maxima in Figures \ref{fig:Iph1} and \ref{fig:Iph2} then occurred around the frequency $\sim 2$ THz, which was used in the experiment of Ref. \cite{Michailow22}. What will change if we take for a typical magnetic field a $K$ times smaller value?

As seen from Eq. (\ref{2ndOrderCurrent-Ww}), if the magnetic field decreases by a factor of $K$, the parameters $\gamma_{\rm edge}/\omega_c$, $T/\hbar\omega_c$ and $H$ increase by a factor of $K$. As we have seen from Figure \ref{fig:Iph2}, a change in temperature has virtually no effect on the absolute value of the function ${\cal J}_{\rm ph,edge}$, increasing the parameter $\gamma_{\rm edge}/\omega_c$ by $K$ times leads to a decrease of ${\cal J}_{\rm ph,edge}$ by $\sim K$ times, while increasing the parameter $H$ by a factor of $K$ increases the function ${\cal J}_{\rm ph,edge}$ by approximately the same value. The normalized chemical potential $\mu/\hbar\omega_c$ can remain unchanged since the electron density can be chosen appropriately, either by choosing a different sample or by applying a voltage to the right part of the antenna, Figure \ref{fig:geom}(b,c), which will now serve as a gate. As a result, the function ${\cal J}_{\rm ph,edge}$ remains approximately the same if the
magnetic field is reduced by a factor of $K$. 

The product of the first and second factors in Eq. (\ref{2ndOrderCurrent-Ww}) is inversely proportional to $B$, therefore, a $K$-fold decrease in the magnetic field should lead to a $K$-fold increase in the photocurrent. In addition, the resonant frequency at which the maximum photocurrent is observed will shift to the region of lower frequencies also by the factor of $K$. The photocurrent estimate of $\sim 2.4$ $\mu$A at the frequency $\sim 2$ THz obtained above for a magnetic field of 4 T is thus converted to the photocurrent of the order of $\sim 10$ $\mu$A at the frequency $\sim 0.5$ THz if the magnetic field is reduced to 1 T, a value which can be achieved with a permanent magnet.

In the estimates given above, it is of course necessary to remember that when the magnetic field decreases by $K$ times, the condition $\omega_c\tau_{\rm edge}\gtrsim \omega_c\tau_p\gg 1$ should not be violated. For example, in samples with a rather moderate mobility $\mu=10^5$ cm$^2$/Vs and in the magnetic field of 1 T, the parameter $\omega_c\tau_p\approx 10$ is still significantly greater than 1.

\section{Discussion and conclusions\label{sec:conclud}}

We have presented a quantum theory of the microwave/THz photoresponse of edge 2D electrons placed in a strong magnetic field. In our work, we considered only conventional semiconductor systems in which electrons have a parabolic energy dispersion.

In the literature, some studies of photocurrents generated by terahertz radiation in a magnetic field were carried out on graphene materials, albeit without an antenna structure. For example, in Refs. \cite{Plank19,Candussio21} such edge photocurrents were experimentally studied in monolayer and bilayer graphene. In these papers, the authors proposed some theoretical arguments, compared their predictions with the experimentally obtained photocurrent, and found a reasonable agreement between their theory and experiment. The absolute value of the theoretically predicted and experimentally measured photocurrent in Refs. \cite{Plank19,Candussio21} was about 
\be 
\frac{I_{ph}}S\sim 0.2 \ \frac{\textrm{nA cm}^2}{\textrm{W}} \textrm{ (graphene)} , 
\ee
where $I_{ph}$ is the photocurrent and $S$ is the intensity of the incident radiation. Let us compare our results obtained for 2D electrons in semiconductor structures with those obtained in Refs. \cite{Plank19,Candussio21} for graphene. To this end, we rewrite the formula (\ref{2ndOrderCurrent-Ww}) in the form
\be 
\frac{I_2^{(0)}}S=
\frac {e^2}{\hbar c}
\left(\frac{E_{\rm ac}}{E_0}\right)^2\frac{eg_sg^2}{\hbar \omega_c}
{\cal J}_{\rm ph,edge} ,
\label{efficienc}
\ee
where $S=cE_0^2/4\pi$ is the intensity (Poynting vector) and $E_0$ is the electric field of the incident electromagnetic wave. The amplification of the electric field in the gap between the two antenna wings $E_{\rm ac}/E_0$ was studied in Ref. \cite{Chen24}. For parameters typical for the IPPE-type experiment \cite{Michailow22} it was found to be $E_{\rm ac}/E_0\sim 40-90$. Taking for this value $E_{\rm ac}/E_0=50$, for the gap width $g=270$ nm \cite{Michailow22}, for the magnetic field $B=4$ T ($\hbar\omega_c=6.89$ meV for the GaAs effective mass) and for the factor ${\cal J}_{\rm ph,edge} $ the value ${\cal J}_{\rm ph,edge} = 2$, we get the photocurrent generation efficiency
\be 
\frac{I_2^{(0)}}S\approx 7.7 \ \frac{\textrm{$\mu$A cm}^2}{\textrm{W}}  \textrm{ (GaAs)} .
\ee
This value is about $38000$ times larger than in graphene \cite{Plank19,Candussio21}. Thus, graphene is not the best material for a detector of the type considered here. 

Why is graphene not suitable for observing the effect discussed here? There are several reasons. Firstly, due to the vanishing effective mass, the cyclotron energy in graphene is huge as compared to conventional semiconductors. As mentioned in Ref. \cite{Plank19}, in the magnetic field of 1 T the cyclotron gap is already 30 meV, so that the energy of the THz photon is much smaller than $\hbar\omega_c$. As seen from Figures \ref{fig:Iph1} and \ref{fig:Iph2}, the photocurrent generation due to the inter-Landau-level transitions is negligibly small at $\omega\ll\omega_c$. In order to bypass the problem, the authors of \cite{Plank19} took into account intra-Landau-level transitions. However, indirect optical transitions within the same Landau level require the simultaneous electron-photon interaction and electron scattering by static impurities or phonons to satisfy the energy and momentum conservation. This significantly reduces the probability of corresponding current generation processes. Secondly, for the same reason -- the large value of $\hbar \omega_c$ -- the Fermi energy also had to be chosen small compared to the cyclotron gap in Ref. \cite{Plank19}. But as can be seen from Fig. \ref{fig:Iph1}, the effect is also very small if $\mu\lesssim \hbar\omega_c$ and increases rapidly when the chemical potential becomes much larger than $\hbar \omega_c$. Thirdly, in the work \cite{Plank19} the entire sample was irradiated. But, as we discussed in Section \ref{sec:intro}, the current generated on one side of the sample is compensated by the current generated on the other side. That is, without the antenna focusing radiation on one side of the sample, Fig. \ref{fig:geom}(b), the generated photocurrent will be significantly less than it could be.

Thus, a photodetector based on 2D edge electrons in a strong magnetic field should be built on conventional semiconductor materials with parabolic electron energy dispersion. Semi-metallic graphene with the linear electron energy dispersion is not suitable for this purpose.

Finally, one more theoretical question requires clarification. We have solved the quantum mechanical problem and obtained a broad asymmetric resonance for the photoresponse, starting from $\omega=\omega_c$ and smoothly decaying up to frequencies of 2, 3, and even $4\omega_c$. This happens because near the edge, quantum transitions between Landau levels $n$ and $n'$ with $|n'-n|>1$ are allowed, see Figure \ref{fig:edgetrans}, and the transition energies can lie between $\omega_c$ and several $\omega_c$. The same result should have been obtained if the problem were solved classically, taking into account the skipping electron orbits near the edge, since, as is known, skipping electrons oscillate with all frequencies from $\omega_c$ to several $\omega_c$. However, in work \cite{Durnev21} by solving the classical Boltzmann equation, a photocurrent spectrum was obtained in the form of a narrow symmetric resonance line with a center at $\omega =\omega_c$ and a width $\sim 1/\tau_p$, see Figure 3 in Ref. \cite{Durnev21}. The question arises why the semiclassical solution of Ref. \cite{Durnev21} contradicts the results of our quantum mechanical solution shown in Figures \ref{fig:Iph1} and \ref{fig:Iph2}. 

The reason for the disagreement is that the boundary conditions used in Ref. \cite{Durnev21} for the distribution function do not properly describe the skipping electron orbits. The authors considered specular reflection of electrons from the sample edge and imposed the following boundary condition on the distribution function:
\be 
f(p_x,p_y,x=0,t)=f(-p_x,p_y,x=0,t).\label{bc-durnev}
\ee
However, this condition is correct and can be used only for electrons with $p_x>0$. If $p_x>0$, it means that the number of electrons moving away from the sample boundary $x=0$ to the right is equal to the number of electrons approaching this boundary from the right. If $p_x<0$, then the condition (\ref{bc-durnev}) does not make sense, therefore the problem posed in work \cite{Durnev21} is not fully defined. This explains why the result obtained in Ref. \cite{Durnev21} is incorrect.

To summarize, we have proposed a method of photocurrent generation by edge electrons in 2D electron systems placed in a perpendicular magnetic field and a suitable geometry of the corresponding microwave/THz detector. We have developed a detailed quantum-mechanical theory which shows that the photocurrent response has a strong and broad resonance at frequencies mainly lying between $\omega=\omega_c$ and several $\omega_c$. We have studied how various physical parameters such as electron density, edge electron scattering rate, temperature, and device dimensions affect the photocurrent response and identified conditions that maximize the photoresponse. We have shown that graphene is unsuitable for building photodetectors based on the skipping 2D edge electrons in a strong magnetic field and that conventional semiconductor materials should be used for this purposes. The proposed detection technique can be used in a wide frequency range covering the microwave to THz frequencies.

\acknowledgments

W.M. thanks Trinity College Cambridge for a Junior Research Fellowship.

\appendix 

\section{Parabolic cylinder functions\label{app:parcylfunc}}

The second-order differential equation 
\be 
\Phi''(x)-\left(a+\frac{x^2}4\right)\Phi(x)=0\label{de1}
\ee
has two independent solutions, the parabolic cylinder functions $U(a,x)$ and $V(a,x)$, Ref. \cite{Abramowitz64}. We need these functions both for positive and negative $x$. Here, we follow the definitions of Ref. \cite{Gil06}. If $x\ge 0$, the functions $U(a,x)$ and $V(a,x)$ are defined in Ref. \cite{Gil06} as in Ref. \cite{Abramowitz64}. If $x<0$, they are defined in Ref. \cite{Gil06} as follows:
\begin{itemize}
\item if $a\le 0$ and $x<0$ then
\be 
U(a,x)=\Gamma\left(\frac 12-a\right)\cos(\pi a)V(a,-x)-\sin(\pi a)U(a,-x),
\ee
\be 
V(a,x)=\sin(\pi a)V(a,-x)+\frac{\cos(\pi a)}{\Gamma\left(\frac 12-a\right)}U(a,-x);
\ee
\item if $a\ge 0$ and $x<0$ then
\be 
U(a,x)=\frac{\pi}{\Gamma\left(\frac 12+a\right)}V(a,-x)-\sin(\pi a)U(a,-x),
\ee
\be 
V(a,x)=\sin(\pi a)V(a,-x)+\frac{\cos^2(\pi a)}{\pi}\Gamma\left(\frac 12+a\right)U(a,-x).
\ee
\end{itemize}
The functions defined this way have the following asymptotic behavior at $x\to\infty$: 
\be 
U(a,x)\sim x^{-a-1/2}e^{-x^2/4}(1+O(x^{-2})), \ \ 
V(a,x)\sim \sqrt{\frac 2\pi}x^{a-1/2}e^{x^2/4}(1+O(x^{-2})),
\ee
that is, the function $U(a,x)$ tends to zero, while the function $V(a,x)$ tends to infinity at large $x$.

\bibliography{}

\end{document}